\documentclass[twocolumn,showpacs,preprintnumbers,amsmath,amssymb,10pt]{revtex4}

\usepackage{graphicx}
\usepackage{dcolumn}
\usepackage{bm}

\begin{document}

\title{Analytical solution of the equation of motion for a rigid domain wall in a magnetic material with perpendicular anisotropy}

\author{M. C. Hickey} \altaffiliation[Electronic mail : ]{hickey@mit.edu}
\affiliation{%
Francis Bitter Magnet Laboratory, Massachusetts Institute of Technology, 150 Albany Street, Cambridge, MA 02139 USA.
}%

\begin{abstract}
This paper reports the solution of the equation of motion for a domain wall
in a magnetic material which exhibits high
magneto-crystalline anisotropy. Starting from the Landau-Lifschitz-Gilbert equation for field-induced motion, we solve the equation to give an analytical expression, which specifies the domain wall position
as a function of time. Taking parameters from a Co/Pt multilayer system, we find good quantitative agreement between calculated
and experimentally determined wall velocities, and show that high field uniform wall motion occurs when wall rigidity is assumed.
\end{abstract}
\maketitle
\begin{figure}[!h]
\includegraphics[width=6.0cm]{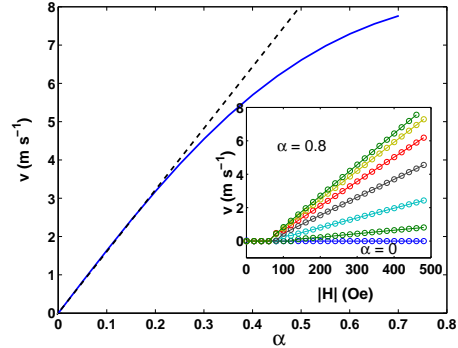}
\caption{(Color Online) Plot of wall velocity at $|H|$ = 500 Oe as a function
Gilbert damping constant. The linear trend (dashed line) corresponds to the precessional regime for small $\alpha$.
The inset shows the field dependent velocity at a range of Gilbert damping parameters. This calculation used magnetic parameters from the Pt/Co(0.5nm)/Pt multilayer system of Metaxas {\it et al.} \cite{metaxas:217208}. }
\label{Velocity_Vs_Field2}
\end{figure}

The area of domain wall spintronics is currently enjoying its heyday, both as a fruitful
discipline for investigating how conduction electrons impart angular momentum onto lattice magnetization spins \cite{bergerpla1973} and
 from the point of view of industrial application. Dynamical studies in domain wall transport \cite{atkinsonnaturemat2003} have led to their use as memory bits \cite{CowburnRussellP.2007,StuartS.P.Parkin04112008} while domain walls also play a central role in magnetic logic devices \cite{D.A.Allwood09092005}. Controlling nano-pillar magnetization with electron current \cite{slonczewskijmmm1996} has been widely demonstrated and forms the basis for magnetic random access memory.\\
Many studies on domain wall motion necessitate a full numerical treatment of the Landau-Lifschitz-Gilbert (LLG) equation
together with a description of the total magnetostatic energy. While the starting descriptions of the magnetostatic energy are well understood, the final numerical simulation often lacks the transparency of a purely analytical treatment. Domain wall motion in Permalloy thin films is richly complicated by a variety of topological structures which can be nearly energetically degenerate. Complications of domain wall distortion under field include the Walker breakdown effect and more generally, oscillatory motion, contraction and expansion of walls which are commensurate with the emission of spin waves. These effects are instabilities and the treatment of the wall as a singular object breaks down as the wall dissipates energy to the lattice. While permalloy is an attractive material from the point of view of low magnetization switching fields and low anisotropy, this type of non-linear behavior is best avoided for reproducible shuttling of domain walls down a patterned magnetic wire.
In this paper, we focus on the description of domain wall motion in a perpendicularly magnetized material (such as a Co/Pt multilayer). We show that, having assumed a rigid wall profile and negligible wall distortion (negligible spatial dependence of wall tilt angle), an analytical solution of the equation of motion of the wall under field comes out, and there are well defined limits where the domain wall motion is robustly linear. The assumption of negligible wall distortion is justified in these materials because the easy axis of the system is always perpendicular to the direction of motion.
We begin with the LLG equation
\begin{eqnarray}
\frac{d {\bf M}}{dt}=\gamma ({\bf M}\times
{\bf H}_{eff})-\frac{\alpha}{M_{s}}({\bf M}\times \dot{\bf M})\label{LLG}
\end{eqnarray}
where $\gamma$ is the gyromagnetic ratio defined as
$\gamma=g\frac{\mu_{B}}{\hbar}$, (g is the electronic g factor,
and $\mu_{B}$ Bohr magnetron) and $\alpha$ is the Gilbert damping. We write the
effective magnetic field in the system as follows :
\begin{eqnarray}
{\bf H}_{eff}=-\frac{1}{\mu_{0}}\frac{\delta E_{d}}{\delta {\bf m}}.\label{Heff}
\end{eqnarray}
E$_{d}$ is the energy density which contains the exchange, uniaxial and
magnetostatic external field energies as described by
equation \ref{Ed}. In spherical coordinates it is written as:
\begin{eqnarray}
E_{d}=A[(\nabla\theta)^{2}+\sin^{2}\theta(\nabla\phi)^{2}]-K\cos^{2}\theta-\mu_{0}{\bf M}\cdot
{\bf H}\label{Ed}
\end{eqnarray}
where K is easy axis anisotropy constant, A is the exchange constant and $\mu_{0}$ is the magnetic permeability of free space while $\theta$ and $\phi$ are the
spherical polar angles of the magnetization.
\begin{eqnarray}
\nabla_{m}=\left( \frac{\partial}{\partial
m},\frac{1}{m}\frac{\partial}{\partial\theta},
\frac{1}{m\sin\theta}\frac{\partial}{\partial\phi} \right).
\label{gradient}
\end{eqnarray}
The magnetization (M=(M$_{x}$,M$_{y}$,M$_{z}$)) can be written in terms of
the spherical polar angles (in a cartesian vector basis) as
${\bf M}=M_{s}(\sin\theta\cos\phi,\sin\theta\sin\phi,\cos\theta)$, where
$\phi=\phi(x,t)$ and $\theta=\theta(x,t)$ are the azimuthal and polar angles, respectively. We can write the time derivative of the
magnetization in the basis vectors of
spherical polar coordinates (e$_{m}$, e$_{\theta}$, e$_{\phi}$).
This is a more convenient coordinate basis, because the magnetic state of the system can be described by
two scalar fields, representing the spherical polar angles, in the above set of equations. Further, only two coupled equations in $\phi$ and $\theta$ are required to describe the magnetostatics and dynamics (see for example, Thiaville {\it et al.} \cite{Thiaville_dynamics}).
Equation \ref{LLG} now reads :
\begin{equation}
\begin{pmatrix}
   \dot{M}_{s} \\
  M_{s}\dot{\theta} \\
  M_{s}\sin\theta\dot{\phi} \\
\end{pmatrix}=\frac{\gamma}{\mu_{0}}
\begin{pmatrix}
  0 \\
  \frac{1}{\sin\theta}\frac{\delta E}{\delta \phi} \\
  -\frac{\delta E}{\delta \theta} \\
\end{pmatrix}
+\frac{\alpha}{M_{s}}
\begin{pmatrix}
  0 \\
  M_{s}^{2}\sin\theta\dot{\phi} \\
  -M_{s}^{2}\dot{\theta}
\end{pmatrix}.
\end{equation}
From this matrix equation, we have a system of two coupled partial differential equations, which
are first order in time.
We can eliminate $\dot{\phi}$ from the system of equations, and we then arrive at the following more simplified
equation describing the time evolution of the magnetization angle $\theta$  ;
\begin{eqnarray}
\dot{\theta}=\frac{1}{M_{s}(1+\alpha^2)}\left[-\frac{\gamma}{\mu_{0}}\frac{1}{\sin\theta}\frac{\delta
E}{\delta\phi}+\frac{\alpha\gamma}{\mu_{0}}\frac{\delta
E}{\delta\theta}\right].
 \label{thetadot}
\end{eqnarray}
We calculate the effective magnetic field (Equation \ref{Heff}) by means of variational calculus, in the following way :
\begin{eqnarray}
\frac{\delta E_d}{\delta\theta}=\frac{\partial
E_d}{\partial\theta}-\frac{d}{dx_{i}}\left(\frac{\partial
E_{d}}{\partial(\frac{\partial\theta}{\partial x_{i}})}\right),
 \label{derivation}
\end{eqnarray}
where repeated indices are summed over and we have a similar equation for the azimuthal angle, $\phi$. We now evaluate these expressions using the definition of the total magnetostatic energy
from Equation \ref{Ed} and substitute these evaluated expressions into Equation \ref{thetadot};
\begin{eqnarray}
\nonumber\dot{\theta}&=&\frac{1}{M_{s}(1+\alpha^2)}[-\frac{\alpha}{\mu_{0}}2A\frac{\sin
2\theta}{\sin\theta}\nabla\theta\cdot\nabla\phi
-\frac{\alpha}{\mu_{0}}2A\sin\theta\nabla^{2}\phi\\\nonumber
&-&\frac{\alpha\gamma}{\mu_{0}}A\sin2\theta(\nabla\phi)^{2}\\&-&(\frac{2\alpha\gamma}{\mu_{0}}K\cos\theta+\alpha\gamma
M_{s}H)\sin\theta+\frac{2\alpha\gamma}{\mu_{0}}A\nabla^{2}\theta].
\label{tetadot2}
\end{eqnarray}
\begin{figure} 
\includegraphics[width=6cm]{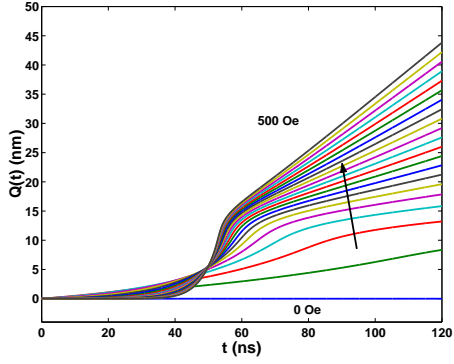}
\caption{(Color Online) The wall position is plotted as a function of time for applied fields between 0 and 500 Oe. The Gilbert damping constant is fixed at $\alpha$ = 0.016. The arrow marks the increasing field.} \label{Qvst}
\end{figure}
We now write down the magnetization of the wall, as a magnetostatic solution, and assume that the wall is rigid and
undergoes no distortion (i.e. $\nabla\phi={\bf 0}$, $\nabla^2\phi=0$). The magnetization for a Bloch wall in a material with perpendicular easy axis anisotropy is taken to be
${\bf M}=M_{s}(0,1/\cosh(\frac{x-Q(t)}{\lambda}),\tanh(\frac{x-Q(t)}{\lambda}))=M_{s}(\sin\theta\cos\phi,\sin\theta\sin\phi,\cos\phi)$,
where x is the central coordinate of the wall magnetization and Q(t) is the position of
the center of the wall.
We use the following parameterization for the magnetization angle ; $\theta$ as $\theta=\cos^{-1}\tanh(\frac{x-Q(t)}{\lambda})$ and insert this definition into the equation of motion given by
Equation \ref{tetadot2}, and arrive at the following first order equation :
\begin{eqnarray*}
-\dot{Q}=\frac{\lambda}{M_{s}(1+\alpha^{2})}[-M_{s}H\alpha\gamma+2\frac{\alpha\gamma}{\mu_{0}}\tanh(\frac{-Q}{\lambda})[-K+\frac{A}{\lambda^{2}}]]
\label{Qdot}
\end{eqnarray*}
This equation can now be integrated, and an implicit solution for the wall position versus time is found to be :
\begin{equation}
(1+\beta u)(u+\beta u^{2})=\beta e^{\frac{2(A+C)}{\lambda}(t+t_{0})},
\end{equation}
where u = e$^{-2(-Q)/\lambda}$ and the constants A and C are defined below in terms of the parameters of the magnetic material and t$_{0}$ is an arbitrary constant.
We can solve this equation above (whose left hand side is cubic in u) to
find the solution in the explicit form Q(t) = F(A,C,t). The result of this inversion is as follows :
\begin{equation}
Q(t) = \frac{\lambda}{2}\ln(y-\frac{2}{3\beta})
\end{equation}
where y is given by the following relation :
\begin{eqnarray*}
y =\frac{\alpha_{1}}{3((-\gamma_{1}-\sqrt{\gamma_{1}^2+4\alpha_{1}^{3}/9})/2)^{\frac{1}{3}}}\\
-((-\gamma_{1}-\sqrt{\gamma_{1}^2+4\alpha_{1}^{3}/9})/2)^{\frac{1}{3}}.
\end{eqnarray*}
The quantities $\alpha_1$ and $\gamma_1$ are given by $(7\beta-4)/3\beta^{3}$ and $(72\beta^{2}-10+27\beta -\beta e^{\frac{2(A+C)}{\lambda}(t+t_{0})})/(27\beta^{3})$, respectively while $\beta = (A-C)/(A+C)$
and we define the constants A and C, as follows ; A = $-(\lambda\alpha\gamma H_{app})/(1+\alpha^2)$, $C = (\lambda2\alpha\gamma/M_s(1+\alpha^2)\mu_0)(-K+A/\lambda^2)$ and we choose the boundary condition dQ/dt(t=0)=0.\\
\begin{figure}[!h]
\includegraphics[width=6.0cm]{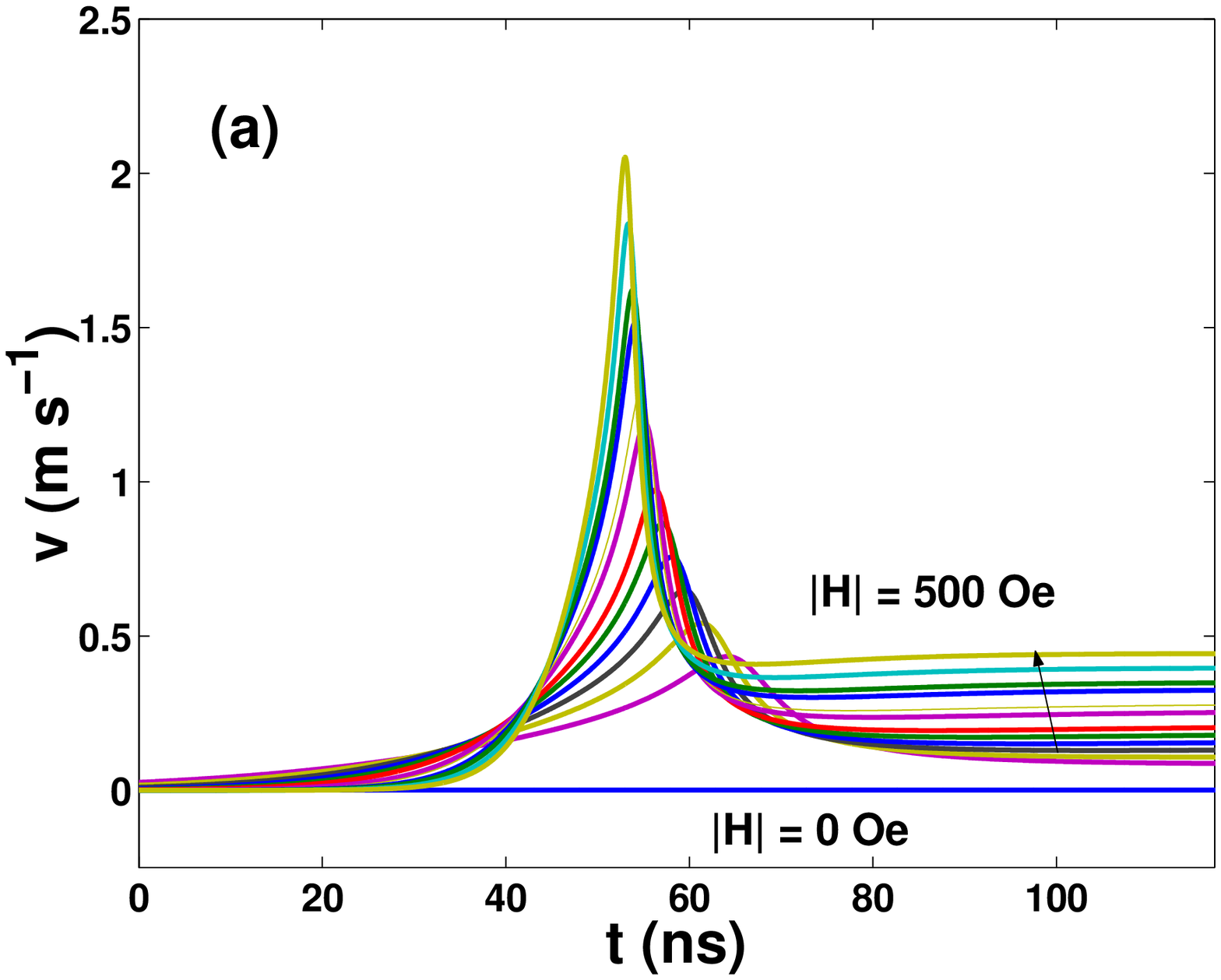}
\includegraphics[width=6.0cm]{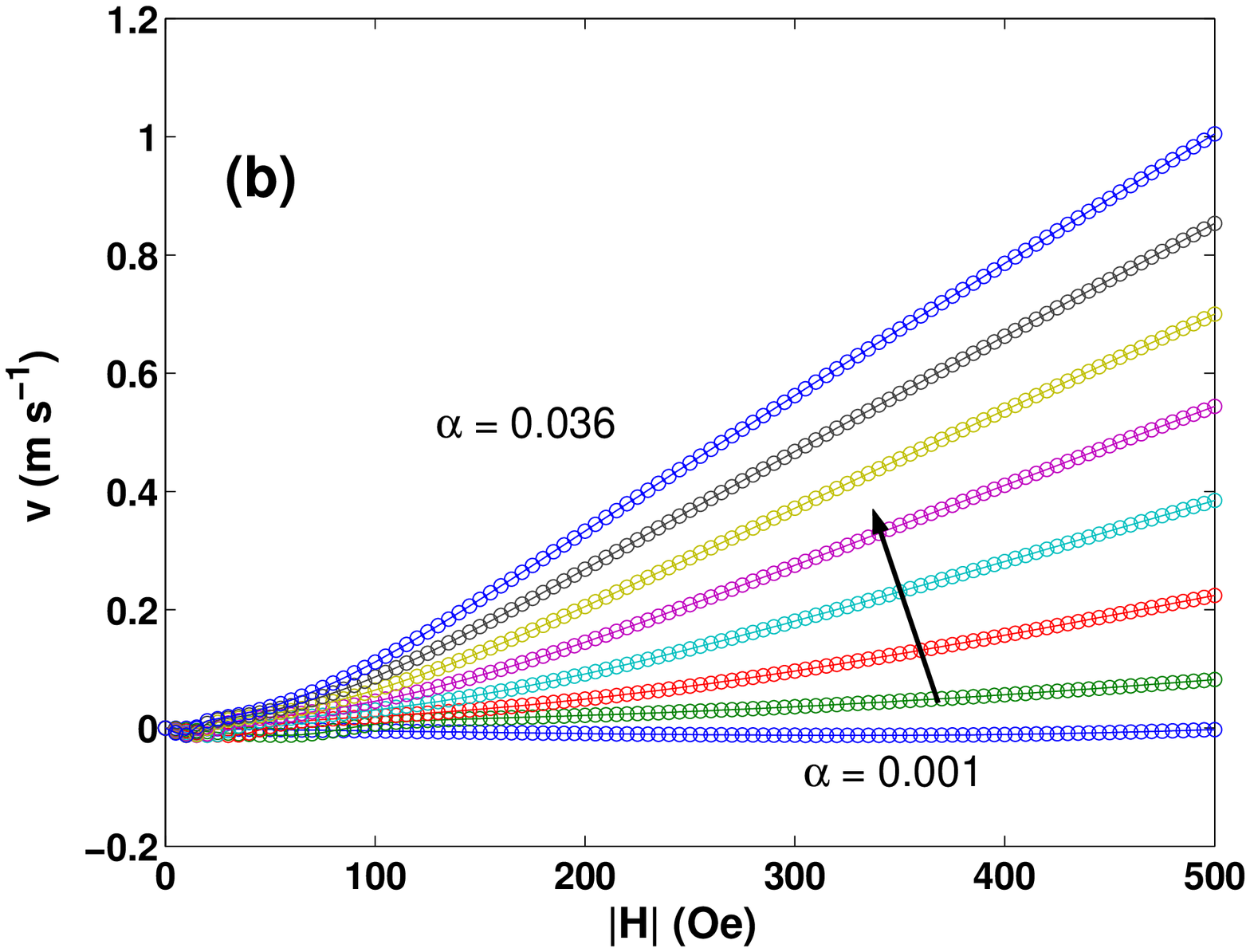}
\caption{(Color Online) (a) Plot of the instantaneous velocity attained by the domain wall under motion by applied field at a fixed Gilbert damping parameter of $\alpha$ = 0.016. The plotted wall velocities here are for applied fields 0 Oe to 500 Oe, the arrow indicates the increase in field magnitude. The flat
region of constant wall velocity is preceded by a critical region. (b) Plot of wall velocity as a function
of field H at differing Gilbert damping constants showing the onset of wall propagation which occurs
when the critical field is reached from saturation.}
\label{Velocity_Vs_Field}
\end{figure}
The results of this analytical model are plotted in Figure \ref{Qvst} and we see two distinct regimes - a non-linear region for t $<$ 60 ns and a linear regime which takes over at timescales greater than 60 ns for all field values.
The values used here for the calculation are taken from a Co/Pt
multi-layer material system \cite{alvarez:09F508} with perpendicular anisotropy, as follows : $\alpha$=0.016,
 $\gamma$=2.2$\times$10$^{5}$A$^{-1}$ ms$^{-1}$,
$\mu_{0}$=4$\pi\times$ 10$^{-7}$ N A$^{-2}$, exchange constant for Co ;
A=3$\times$ 10$^{-11}$J m$^{-1}$, M$_{s}$=1.5 MA m$^{-1}$, K(=K$_{eff}$)=0.3$\times
$10$^{6}$ Jm$^{-3}$ and $\lambda\sim\sqrt{A/K}$=10 nm. Note that the perpendicular anisotropy constant here K
is an effective anisotropy constant which takes into account the effect of the thin film demagnetization field.
Using these materials parameters, the dynamic wall velocity (v=dQ/dt)) versus times at various applied fields (from 0 to 500 Oe) is shown in Figure \ref{Velocity_Vs_Field} (a) and this gives
steady state wall velocities in the region 0-0.5 ms$^{-1}$. The field direction is chosen so that reverse saturation of the magnetization occurs as the wall moves in the positive x direction. The steady-state (t $>$ 60 ns) wall velocity is plotted in Figure \ref{Velocity_Vs_Field} (b) as a function of applied fields at differing Gilbert damping parameters. These results show that the wall begins to move once a critical field is reached and that the wall velocity has an exponential dependence on field. Further, we plot the wall velocity in the steady-state regime at an applied field of 500 Oe against the Gilbert damping parameter $\alpha$, as shown in Figure \ref{Velocity_Vs_Field2}. Here we find a linear relationship for small $\alpha$ which corresponds to the models developed by Slonczewski \cite{J.C.Slonczewski1972} and others \cite{schryer:5406,malozemoff}, whereby one takes the precessional regime of steady-state wall translation (post Walker breakdown) and writes the wall velocity as : v = $\frac{\gamma\lambda}{\alpha+\alpha^{-1}}H \simeq \gamma\lambda\alpha H,$
and this linear expansion is valid for small $\alpha$. For $\alpha$ = 0.3 and at $|H|=500$ Oe, we have a wall velocity of
$\sim$ 5 ms$^{-1}$. This is in reasonable agreement with recently published results \cite{metaxas:217208} on field driven walls in Pt/Co(0.5 nm)/Pt thin film systems. That work reported experimental wall velocities of $\sim$ 8-10 ms$^{-1}$ at 500 Oe with a Gilbert damping constant of about 0.3, having established anisotropy energy density, exchange stiffness and saturated magnetization all identical to that which we have used to parameterize our analytical model, the results of which are plotted in Figure \ref{Velocity_Vs_Field2} and its inset.

This correspondence arises in the linear regime, where the wall translates uniformly and the models neglect pinning
due to defects. The linear regime occurs after Walker breakdown and in the limit of a perfect wire and corresponds to the precessional regime.\\
In conclusion, we have calculated an analytical solution of the equation of motion for a undistorted domain wall in a perpendicularly magnetized material. This solution is constructed using first principles arguments from energy minimum considerations and the trajectories of the wall are completely specified by material parameters. Under the assumption of wall rigidity, we have linear wall translation above a critical threshold where the wall position is
exponentially dependent upon time. The values for wall velocities in the linear regime are in good agreement with previous experiments
on field-driven walls in Pt/Co(0.5 nm)/Pt thin films, and the wall velocity is linearly dependent upon Gilbert damping
corresponding to precessional motion for small Gilbert damping constant.

\begin{acknowledgments}
The author is grateful to Lara San Emeterio-Alvarez for fruitful discussions.
The author would like to thank EPSRC for funding via the Spin@RT consortium and 
the US-UK Fulbright Commission for financial support.
\end{acknowledgments}


\end{document}